\documentclass[useAMS,letterpaper]{mn2e}
 \usepackage[totalwidth=480pt,totalheight=680pt]{geometry}
 \usepackage{graphicx}

\def\eq{\begin{equation}}
\def\en{\end{equation}}

\def\etal{{\it et al.}}
\def\ie{{\it i.e.}}

\def\eg{{\it e.g.}}
\def\aj{{\it A.J.}}
\def\apj{{\it Ap.J.}}

\def\apjs{{\it Ap.J. Suppl.}}
\def\aap{{\it A\&A}}

\def\mnras{{\it MNRAS}}
\def\P3hat{{\mathaccent 94 P}_3}

 \title[Another Look at B1918+19] 
   {Drifting, moding, and nulling: another look at pulsar B1918+19}
 \author[Joanna M. Rankin, Geoffrey A.E. Wright, and Andrew M. Brown]{Joanna M. Rankin,$^1$\thanks{Joanna.Rankin@uvm.edu} Geoffrey A.E. Wright,$^2$\thanks{G.Wright@sussex.ac.uk}  and Andrew M. Brown$^3$\thanks{Current address: Udacity Corp., Palo Alto, CA; email: andb87@gmail.com}\\
   $^1$Physics Department, University of Vermont, Burlington, VT\\
   $^2$Astronomy Centre, University of Sussex, Falmer, Brighton, UK\\
   $^3$Physics Dept., Massachusetts Institute of Technology, Cambridge, MA}

 \pagerange{\pageref{firstpage}--\pageref{lastpage}} \pubyear{2008}

\def\LaTeX{L\kern-.36em\raise.3ex\hbox{a}\kern-.15em
    T\kern-.1667em\lower.7ex\hbox{E}\kern-.125emX}

\begin{document}

\label{firstpage}

\maketitle
  
\begin{abstract} 
Arecibo observations of the conal triple pulsar B1918+19 at 0.327- and 1.4-GHz are used 
to analyse its subpulse behaviour in detail.  We confirm the presence of three distinct drift modes 
(A,B,C) plus a disordered mode (N) and show that they follow one another in specific cycles.  
Interpreting the pulsar's profile as resulting from a sightline traverse which cuts across an 
outer cone and tangentially grazes an inner cone, we demonstrate that the phase modulation 
of the inner cone is locked to the amplitude modulation of the outer cone in all the drift modes.  
The 9\% nulls are found to be largely confined to the dominant B and N modes, and, in the 
N mode, create alternating bunches of nulls and emission in a quasi-periodic manner with 
an averaged fluctuation rate of about 12 rotation periods ($P_1$).  We explore the assumption 
that the apparent drift is the first alias of a faster drift of subbeams equally spaced around the cones. This is shown to imply that all modes A, B and C have a common circulation time of 12 $P_1$ and differ only in the number of subbeams. This timescale is on the same order as predicted 
by the classic {\bf E}$\times${\bf B} drift and also coincides with the N-mode modulation.  We 
therefore arrive at a picture where the circulation speed remains roughly invariant while 
the subbeams progressively diminish in number from modes A to B to C, and are then 
re-established during the N mode. We suggest that aliasing combined with subbeam loss 
may be responsible for apparently dramatic changes in drift rates in other pulsars.  

\end{abstract}


\section{Introduction}

Pulsar B1918+19 is a typical `slow' radio pulsar with a stellar-rotation period 
($P_1$) of 0.82 s and a spin-down age of 15 Myr. It was one of three apparently 
conal pulsars with three-peaked profiles whose pulse-sequence (hereafter PS) 
behaviour was studied at 430 MHz by Hankins \& Wolszczan (1987; henceforth HW). 
These authors identified four distinct 
modes of emission labelled A, B, C, and N --- a nomenclature that will also be 
used here for continuity.  The first three modes exhibited drifting subpulses 
which, together with the fourth ``disordered'' mode and embedded null sequences, were found to interact in specific sequences.

In this paper we exploit the improved signal-to-noise ratio (S/N) of our observations 
to analyse this pulsar's drift modulation and nulls more thoroughly. A full statistical 
assessment is given for each mode and type of modal sequence.  Recent studies of 
other similar pulsars have found that nulling is sometimes confined to a specific mode 
of emission (Redman \etal\ 2005) or found to have a periodic nature (Rankin \& Wright 
2007; Herfindal \& Rankin 2007, 2009) and we establish here to what degree this is 
true of B1918+19.

To place our results in a physical, or at least geometric, framework, we consider the 
context of a rotating subbeam-carousel system (RS). This has often proved successful 
in describing drifting subpulses in conal pulsars (\eg, Deshpande \& Rankin 1999, 2001; 
hereafter DR; Battacharyya \etal\ 2007), and it is now interesting to explore whether the 
more complex multiple discrete mode patterns of B1918+19 together with their `sequence 
rules' can be made  compatible with such a model.  

In \S 2 we describe our observations, and in \S 3 we present what can be inferred 
about the pulsar's basic emission geometry from its profile and polarization.  \S 4 
introduces the multiple features in its fluctuation spectra, and \S 5 discusses the 
profiles and phase patterns of the several modes.  \S 6 gives an analysis of nulling 
statistics, and  \S 7 discusses mode sequencing and transitions.  In \S 8 we discuss 
the possibility of aliasing, and \S 9 provides an interpretation of the pulsar's three 
drift modes in terms of the rotating carousel-beam model, and this  is checked 
against the observationally-derived geometry of the pulsar in \S10. \S 11 summarises 
the results and \S 12 draws conclusions.


\section{Observations}
The observations were carried out using the 305-meter Arecibo Telescope in Puerto Rico.  
All of the observations used the upgraded instrument with its Gregorian feed system, 
327-MHz (P band) or 1400-MHz (L band) receivers, and Wideband Arecibo Pulsar 
Processors (WAPP\footnote{http://www.naic.edu/$\sim$wapp}).  The ACFs and CCFs 
of the channel voltages produced by receivers connected to orthogonal linearly (circularly, 
after 2004 October 11 at P band) polarized feeds were 3-level sampled.  Upon Fourier 
transforming, some 64 or more channels were synthesized across a 25-MHz bandpass 
with about a milliperiod sampling time.  Each of the Stokes parameters was corrected 
for interstellar Faraday rotation, various instrumental polarization effects, and dispersion.  
The date, resolution, and the length of the observations are listed in Table~\ref{Table1}.  

\begin{table}
\begin{center}
\caption{Band, resolution, and length of Arecibo observations.}
 \begin{tabular}{cccccc}
 \hline
 \hline
Band & MJD & Resolution & Length\\
(GHz) &    & (degrees/sample)  & (pulses)\\
 \hline
 1.1-1.7 & 52735 & 0.22 & 1023 \\
 0.327 & 52942 & 0.22 & 1395 \\
 0.327 & 53778 & 0.39 & 3946 \\
 1.1-1.7 & 54541 & 0.36 & 731 \\ 
\hline
\end{tabular}
\end{center}
\scriptsize
The first two observations were seriously corrupted by interference.
\normalsize
\label{Table1}
\end{table}

\begin{figure}
\begin{center}
\includegraphics[width=80mm, angle=0.]{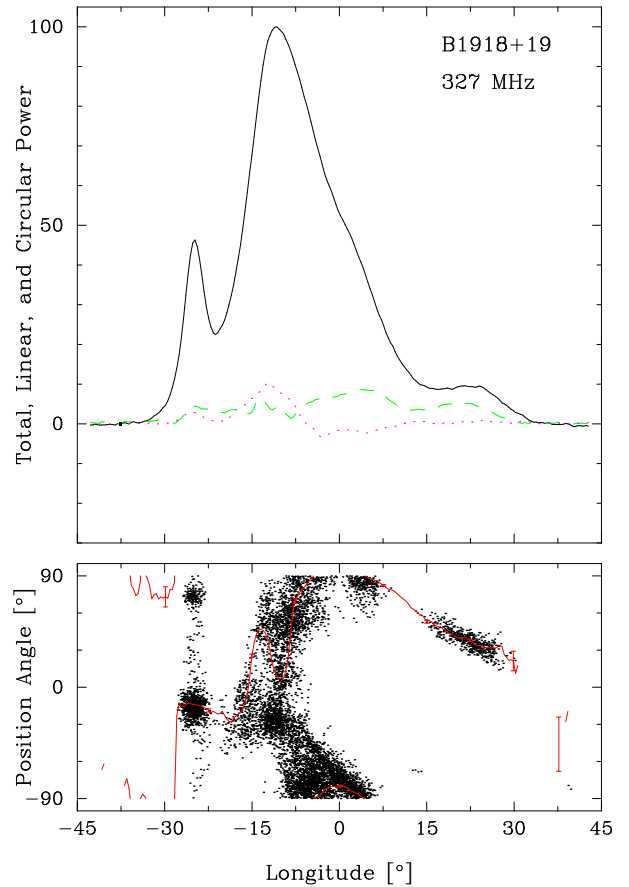}
\caption{Average profile of B1918+19 computed from the long 327-MHz 
observation.  Notice the leading and compound primary components and 
the relatively weak trailing component.  The top panel gives the the total 
intensity (Stokes $I$; solid curve), the total linear (Stokes $L$ (=$\sqrt{Q^2+U^2}$); 
dashed), and the circular polarization (Stokes $V$; dotted).  The lower 
panel PPA (=${1\over 2}\tan^{-1}(U/Q)$) histogram corresponds to those samples 
having PPA errors smaller than 10\degr\ with the average PPA traverse 
overplotted.  The longitude origin is taken about halfway between the 
outer conal component pair.}  
\label{Fig1}
\end{center}
\end{figure}

\begin{figure}
\begin{center}
\begin{tabular}{@{}l@{}l}
{\mbox{\includegraphics[width=3.8cm,height=20cm,angle=0]{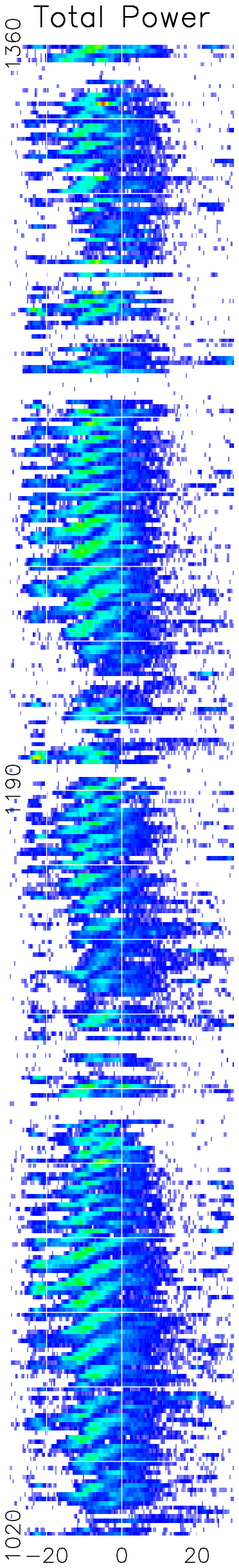}}} & \ \ \ \ \ 
{\mbox{\includegraphics[width=3.8cm,height=20cm,angle=0]{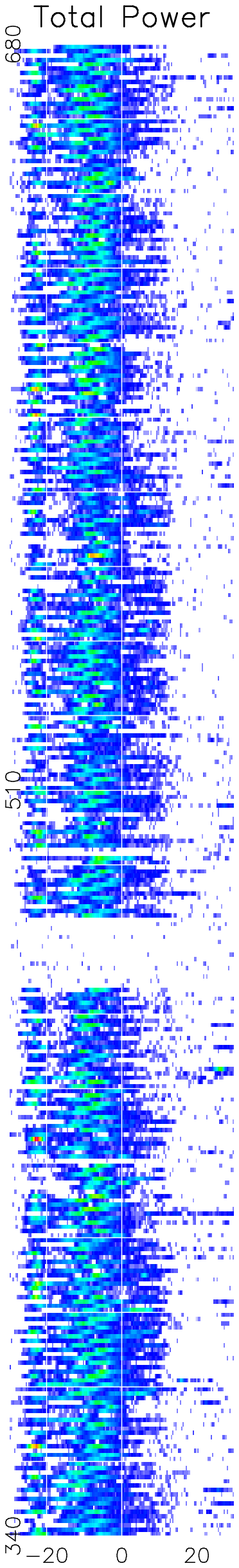}}}\\
\end{tabular}
\caption{Two single pulse plots, each of 340 consecutive pulses from the same observation 
of B1918+19, separated by about 1000 pulses. Left: several sequences of mode A transforming 
into mode B interposed by the null-rich irregular N mode. Note the nulls creating an approximately
 85-pulse periodicity.  Right: an extended sequence of the fast mode C interrupted by an extended 
 null.  Longitude scale in degrees.}
\label{Fig1a}
\end{center}
\end{figure}

\section{Profile Geometry}
The B1918+19 polarization profile and PPA histogram, shown in Figure~\ref{Fig1}, 
consists essentially of four components, a distinct leading component, a central 
 ``main'' component exhibiting a clear shoulder, plus a weak trailing component 
some 35\degr\ of longitude after the peak.  Its profile has therefore been placed 
in the relatively uncommon conal quadruple (cQ) category (Rankin 1993; ET VI), 
and we now know that some other such stars also have highly asymmetric profiles 
(\eg, J1819+1305; see Rankin \& Wright 2008). The available profiles of B1918+19 
(\ie, Gould \& Lyne 1998; Hankins \& Rankin 2008) suggest that it becomes more 
symmetric at higher frequencies, but no published profile at a frequency higher 
than 1.6-GHz seems to exist.  The relative intensity of the first component at meter 
wavelengths also seems to vary considerably, but as we will see below, this is
probably due to the specific population of modes present in the average.  

The aggregate linear polarization $L$ never exceeds about 10\% in Fig.~\ref{Fig1}, 
but the geometric polarization angle (hereafter PPA) sweep rate can be reliably 
traced once the effects of two modal ``flips'' are resolved.  The PPA then makes a 
total traverse of about --120\degr, and the central rate $R$ is some --3.2\degr/\degr.  
Despite the then limited information, B1918+19's basic emission geometry seems 
to have been computed almost correctly in ET VI:  its respective inner and outer 
cones have half-power, 1-GHz widths of 20.5 and 51.3\degr, respectively.  Its magnetic 
latitude $\alpha$ and sightline impact angle $\beta$ are then 14\degr\ and --4.3\degr\ 
for a poleward traverse across its emission cones (or 11\degr\ and +3.4\degr\ for 
an equatorward traverse, although this model fits the inner/outer conal dimensions 
less successfully).  In either case, the sightline encounters the inner cone tangentially 
(with $\beta/\rho \sim$0.9, where $\rho$ is the outer half-power conal radius) and the 
outer cone much more centrally (--0.65/0.52)---confirming the earlier qualitative 
interpretation of the star's double-conal emission geometry.  Moreover, the 
prominent polarization-modal activity on the profile edges seems to confirm that we 
are observing emission from the full widths of the cones.  Note, though, that the PPA 
distribution in the leading half of the central component consists of a set of ``patches'' 
rather than two PPA tracks.


\section{Pulse-Sequence Modulation}
The two extracts from our longer 327-MHz observation presented in Figure~\ref{Fig1a} 
show the essential features of B1918+19's single pulse behaviour. Using the 
nomenclature of HW, the 340 pulses on the left show several examples of a 
relatively slow drift pattern (A) changing rapidly into a faster mode (B), which 
is then  followed by an irregular null-rich sequence (N). The same number of 
pulses in the right panel show a typical long sequence of the fastest mode (C) 
interrupted by a relatively long null. The exceptional S/N of the whole observation 
enables us to analyse these basic features in some detail.

\begin{figure}
\begin{center}
\includegraphics[height =80mm, angle=-90.]{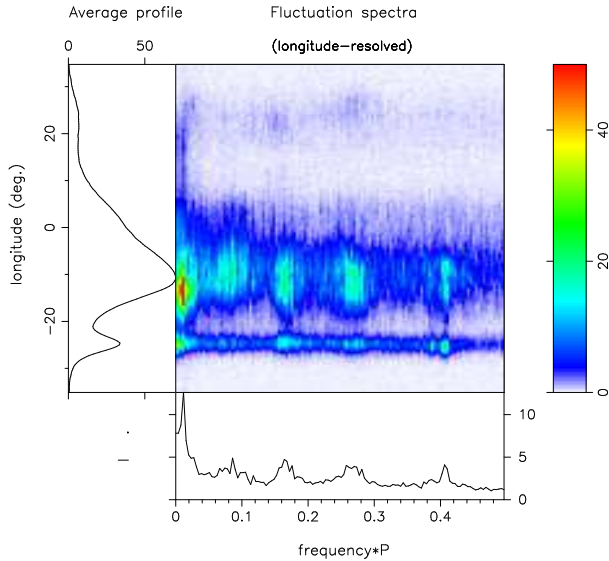}
\caption{Longitude-resolved fluctuation-power spectra for the 327-MHz 
observation of B1918+19 on MJD 53967 (main panel, per the colour scale 
at the right), plotted against the total power profile (lefthand panel) with the 
integral spectrum in the bottom panel.  Five different features are seen in 
these spectra:  The three highest frequency features correspond to the A, 
B and C modes identified by HW with $P_3$ values of 5.8, 3.9 and 2.5 
cycles/$P_1$.  In addition two lower frequency periodicities are present, a 
well defined one corresponding to a $P_3$ value of some 85 $P_1$ and 
a broad response at about 12 $P_1$.  Note that the 2.5-$P_1$, 3.9-$P_1$ 
and 5.8-$P_1$ features all modulate outer regions of the outer cone to some 
degree although the  2.5-$P_1$ feature is not detected in the trailing component.  
Finally, the low frequency feature modulates the entire profile, whereas the 
12-$P_1$ modulation affects the central inner-cone region only.}  
\label{Fig2}
\end{center}
\end{figure}

\begin{figure}
\begin{center}
\includegraphics[height =80mm, angle=-90.]{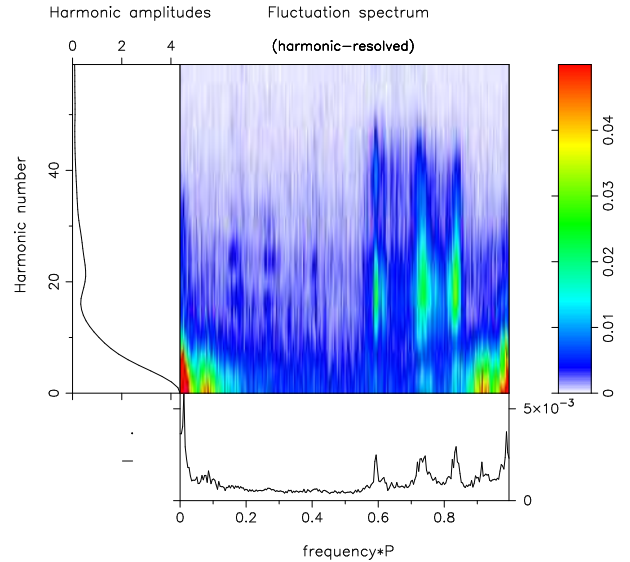}
\caption{Harmonic resolved fluctuation spectrum for the same observations 
as in Fig.~\ref{Fig2} (main panel, per the colour scale at the right).  Rather 
dramatically, this shows that HW's three features are all phase-modulated.  The low frequency response, by contrast, is a mixture of phase and amplitude modulation, and 
the 12-$P_1$ response appears to represent mostly amplitude modulation.  
This distinction can also be seen in the main panel where the low frequency 
modulation is carried by smaller-numbered Fourier harmonics than HW's 
drift modulation, which seem to peak about number 18 or so.  In the central panel the amplitude scale has been compressed by a factor of 3 to accentuate  
the three phase-modulated periodicities.}  
\label{Fig3}
\end{center}
\end{figure}

The longitude- and harmonic-resolved fluctuation spectra (hereafter lrf and hrf) 
of the entire 327-MHz observation are given in Figures~\ref{Fig2} and 
\ref{Fig3}. In the lrf spectra four clear features are visible, in addition to a 
prominent low-frequency feature.  The periodicities corresponding to the A, B, \
and C modes, which represent the vertical spacing of the driftbands and are conventionally denoted as $P_3$, are very evident, with centroid values only 
slightly different from those of HW.  The A-mode has a $P_3$ of 6.06 $P_1$ 
(=1/0.165 c/$P_1$; 5.9 $P_1$ in HW); the B-mode a $P_3$ of 3.80 $P_1$ 
(=1/0.263 c/$P_1$; 3.85 $P_1$ in HW); and the C-mode a $P_3$ of 2.45 $P_1$ 
(=1/0.408 c/$P_1$; 2.5 $P_1$ in HW).  Substantial scatter about these values 
is visible in Fig.~\ref{Fig2} (especially in the B-mode), but we will show below 
that they are often nearly coherent in particular sections of the PS.  Note also 
that the periodicities corresponding to these three modes are found not only 
in the central component of the integrated profile, but also in the outer 
component pair.  Significantly, the A and B modes modulate both the leading 
and trailing components, although on the trailing edges at differing longitudes. 
The C mode modulation is also prominent in the leading component but invisible 
in the trailing one, where the intensity of this mode is anyway very 
weak.\footnote{Weltevrede \etal\ (2006, 2007) were unable to find modulation in 
this pulsar at 21 cms; whereas, at 92 cm they detected two drift features.  See their 
fig. A12.}  




The A-C features also appear prominently in the hrf spectrum of Fig.~\ref{Fig3}, 
but here almost exclusively at fluctuation frequencies between about 0.6 and 
0.85 c/$P_1$.  The main panel shows clearly that these periodicities represent 
phase modulations corresponding to sets of drifting subpulses separated by 
$P_2\sim$20\degr (=360\degr/18) 
 and their presence at frequencies over 0.5 indicate that the \textit{apparent} 
 drift is toward increasing longitudes.  In itself the hrf spectrum cannot determine 
 the \textit{true} subpulse drift, i.e whether or not the observed drift pattern is aliased 
 (in which case the intrinsic drift would be toward decreasing longitudes). This 
 point is further discussed below.

This new periodicity may well arise within the irregular null-rich N-mode, 
which is known to have a peak intensity at a slightly later longitude than the other 
modes and to be characterised by a weak first component.  This will be investigated 
in a later section, but it can be seen in the left-hand PS of Fig.~\ref{Fig1a} 
that intensity fluctuations on roughly this timescale occur between the A-B modal 
sequences. It is interesting to note---not for the first time in pulsars with multiple 
drift patterns---that the periodicities of this pulsar appear to have ``harmonic'' 
properties: the new periodicity (12.0 $P_1$) is practically twice that of the A-mode 
(6.06 $P_1$), which in turn is about three-halves of the mean value of the B-mode 
(3.80 $P_1$) and five halves of the C-mode periodicity (2.45 $P_1$). 

\begin{figure}
\begin{center}
\includegraphics[height=99mm,width=80mm,angle=0.]{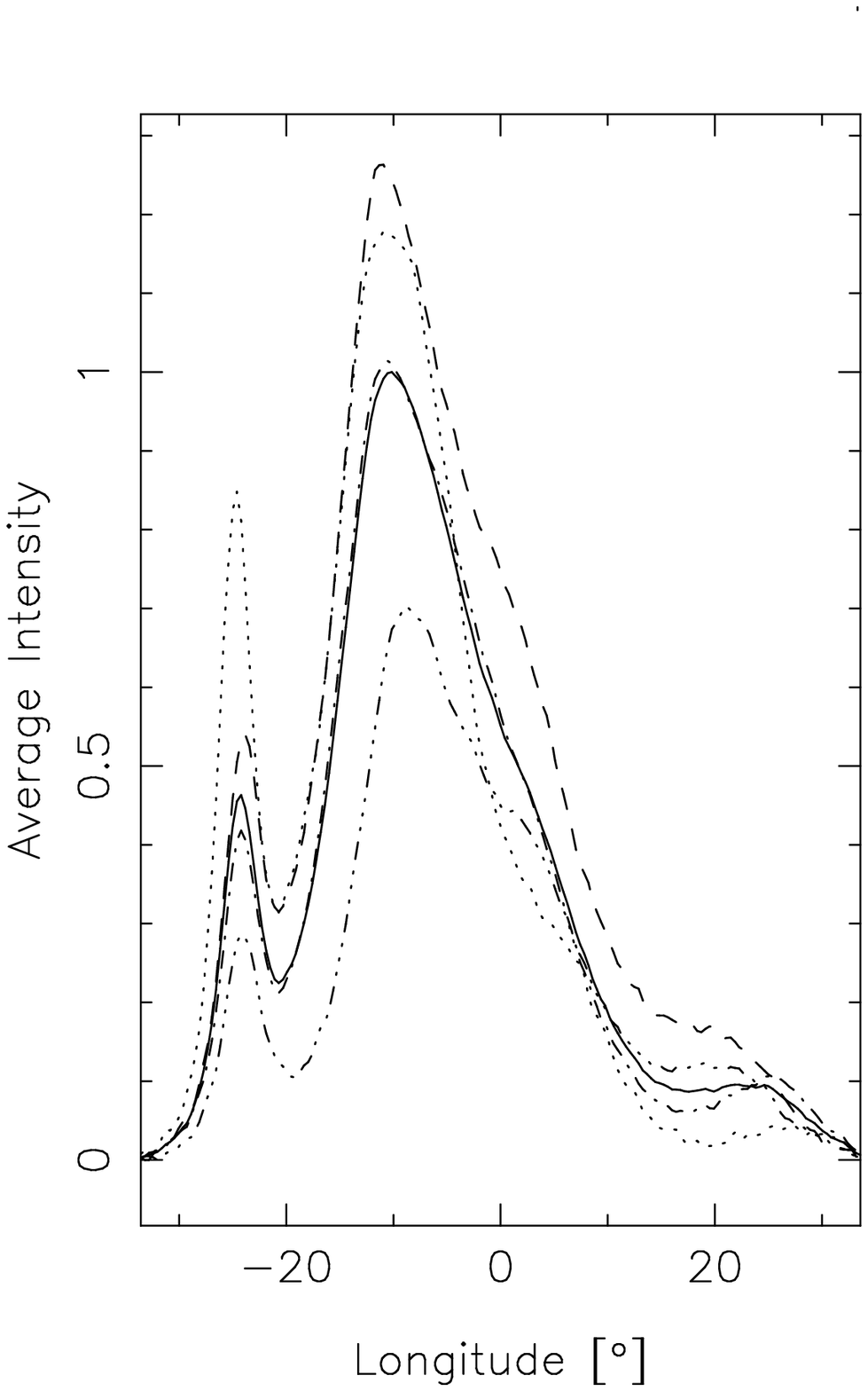}
\includegraphics[height=99mm,width=80mm,angle=0.]{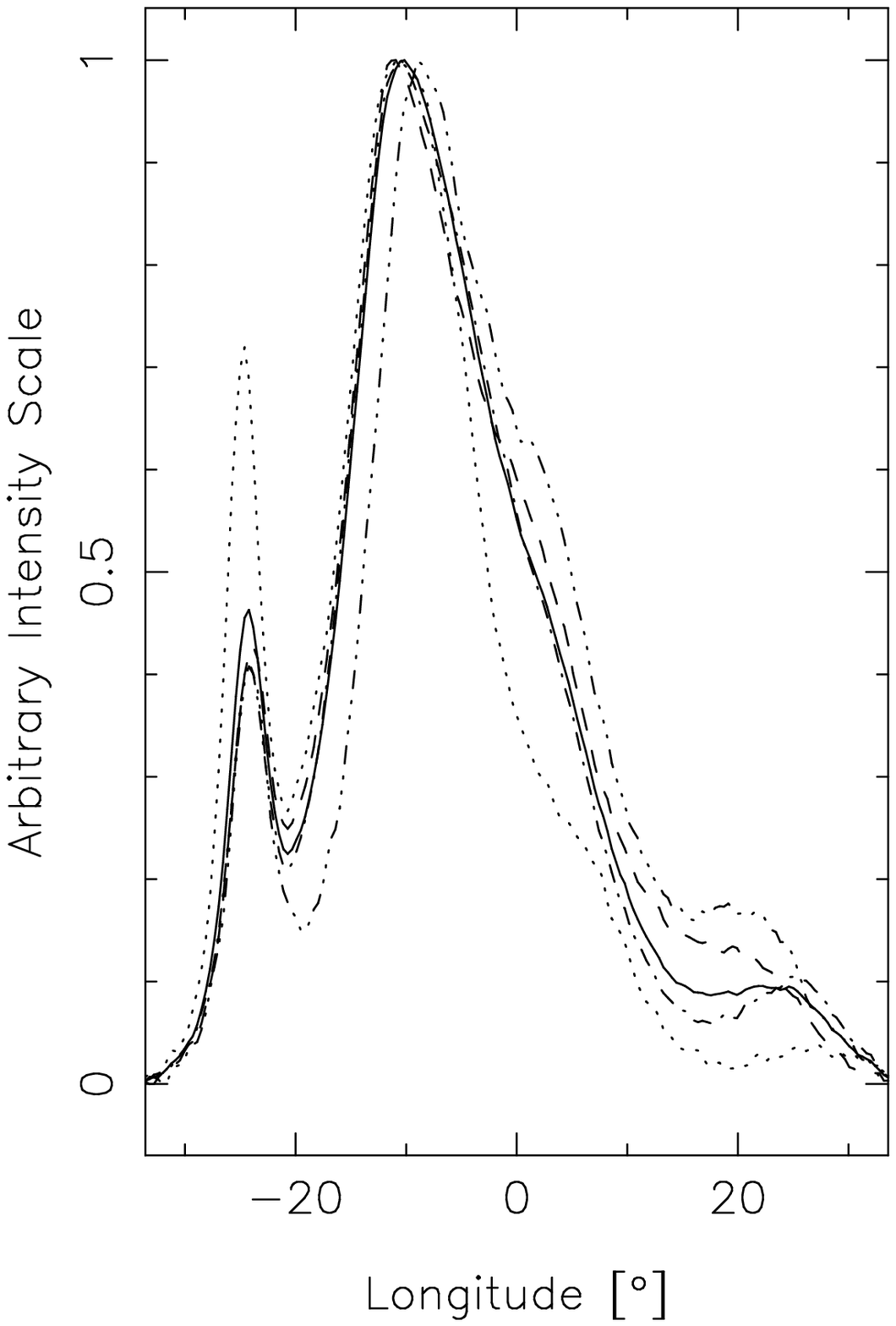}
\caption{Partial profiles for the several modes as well as the average profile: 
(top left) normalized to their relative intensities; (top right) leading component 
detail; and (bottom) scaled to the peak intensities.  Total average profile (solid 
line) as well as the A (dashed), B (dash-dotted), C (dotted), and N (dash-triple 
dot line) modes.  Notice that the leading component of the C mode is significantly 
larger than in any other mode, whereas its trailing component is much smaller.  
It also peaks about 1\degr\ earlier than the A mode, while the B mode peaks 
somewhere between.  
Similarly, the normalized profiles (bottom) clearly show the shift of the peak of 
the N mode as well as the curtailed trailing edge of the C mode.}
\label{Fig4}
\end{center}
\end{figure}

\begin{figure}
\begin{center}
\includegraphics[height=99mm,width=80mm,angle=-0.]{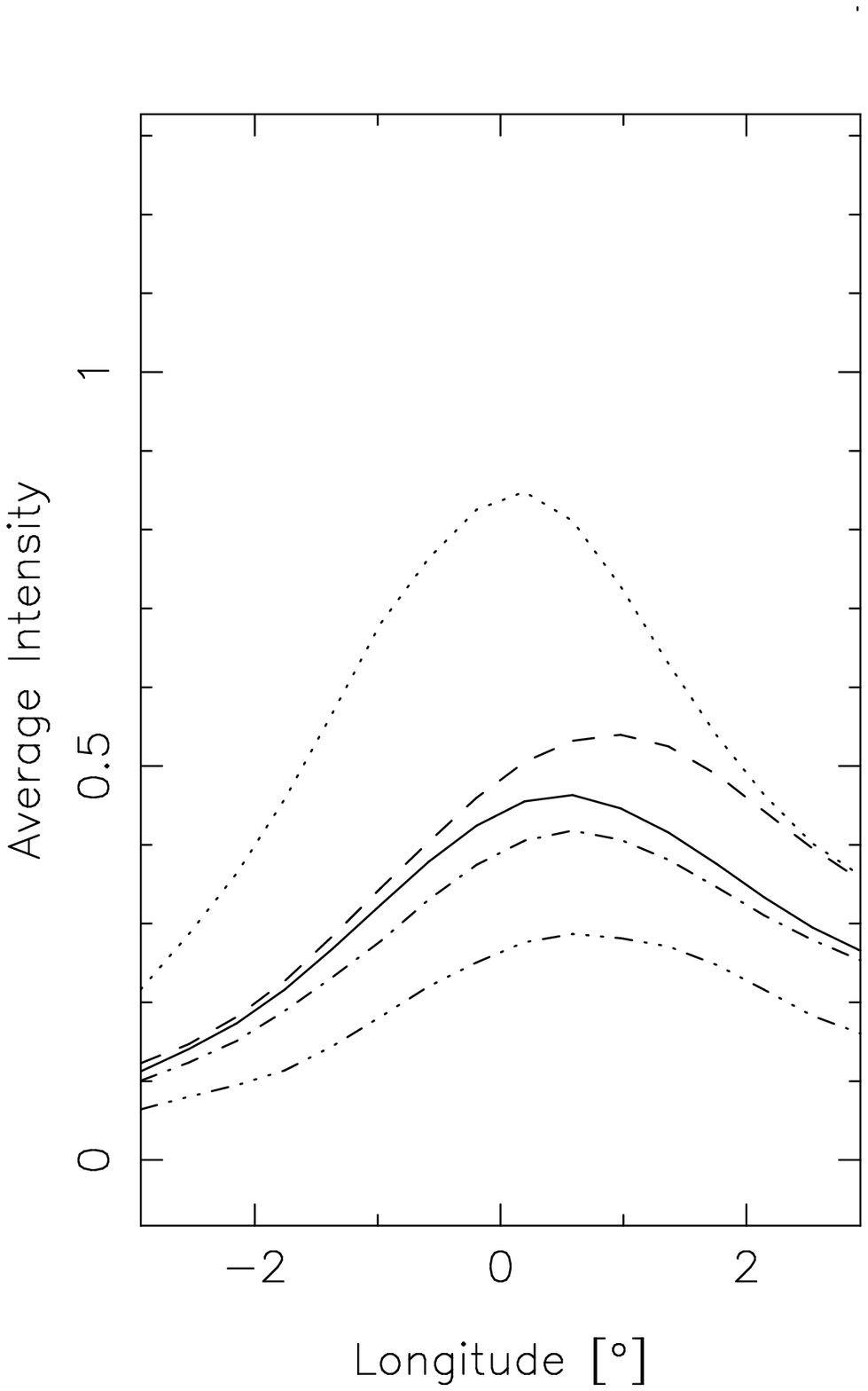}
\end{center}
\end{figure}

Finally, the low-frequency feature in the two spectra---mentioned in passing by HW 
and attributed to ``deep nulls''---has its peak before the profile peak and is prominent 
across nearly the entire profile.  A striking example of how this feature is is generated 
in the PS can be seen in the left of Fig.~\ref{Fig1a}, where the 340 pulses are divided 
into 4 roughly equal sections by the intrusion of the N mode. However, in the PS as a 
whole this represents more of a characteristic timescale rather than a periodicity.  In 
the hrf spectrum (Fig.~\ref{Fig3}) the feature interestingly reveals itself to be an 
admixture of amplitude and phase modulation---the latter counter to that of the A-C 
modes and focussed on a narrow band of longitude preceding the profile peak by 
about 3$\degr$.  (NB, a similar feature was found in J1819+1305 in an earlier paper.)  

As suggested by Fig.~\ref{Fig1a}, these last two periodicities (12-$P_1$ and 85-$P_1$) 
do appear to be null related.  A number of the nulling pulsars exhibit such low frequency 
fluctuation-spectral features, and when the PS modulation is quelled by substituting a 
scaled-down profile for the pulses and zeroes for the nulls, the periodicities often persist 
(Herfindal \& Rankin 2007, 2009).  The latter study shows that this is so for B1918+19 as 
well; three different periodicities appear to be present in the pulsar's null sequence: a 
seemingly harmonically-related pair at 85$\pm$14 and 43$\pm$4 $P_1$ as well as the 
roughly 12-$P_1$ one.

\section{Distinguishing the Modes}
Using colour displays such as those in Fig.~\ref{Fig1a}, we inspected the 
entire long 327-MHz PS to identify sections exhibiting the A, B and C modes, usually 
on the basis of their characteristic periodicity.  Mode N was identified by the absence 
of a clear drift, often accompanied by weakness in the first component.  


By far the most common mode was B (1896 pulses or 48\%), followed by N (847 
pulses or 21.5\%), A (662 pulses or 17\%), and C (541 pulses or 14\%).  However, 
these statistics disguise the considerable discrepancies in the typical sequence 
lengths of the modes:  while all C-mode pulses occurred in just 4 long null-free 
sequences, averaging 135 pulses in length, the 36 N-mode PSs averaged only 
23.5 pulses, and the A- and B-mode PSs averaged 35 and 53 pulses, respectively.

By segregating the pulses appropriately a partial-average profile was computed 
for each mode. Figure~\ref{Fig4} (top left) shows the profiles weighted according  
to their relative average intensities. Note that the average intensities of the A- and 
C-mode pulses are clearly stronger than those of the B and N modes. This differs 
slightly from the conclusion of HW, who found, after selecting 300 bright pulses 
from each mode (at 430 MHz), that the average intensities of all the modes were 
more comparable. Their result can be reconciled with ours if, as seems likely, the
weaker pulses (including nulls) tend to occur in the B or N modes.  Particularly 
striking in both ours and HW's results is the enhanced intensity of the leading 
component in the C mode, which is about double its average strength in the other 
modes. The first component peak appears about 1\degr\ earlier in the C mode than 
in the A mode, with those of B and N modes falling in between (top right panel). 


\begin{figure*}
\begin{center}
\begin{tabular}{@{}l@{}l}
{\mbox{\includegraphics[width=7.2cm,angle=-90]{PQmod_fold_B1918+19.53778_A_2421-2464.ps}}}\ \ \ \ \ \ \ \ \ &
{\mbox{\includegraphics[width=7.2cm,angle=-90]{PQmod_fold_B1918+19.53778p_B_2600-2800.ps}}}\\

\end{tabular}
\end{center}
\end{figure*}

\begin{figure}
\begin{center}
\includegraphics[height=80mm,angle=-90.]{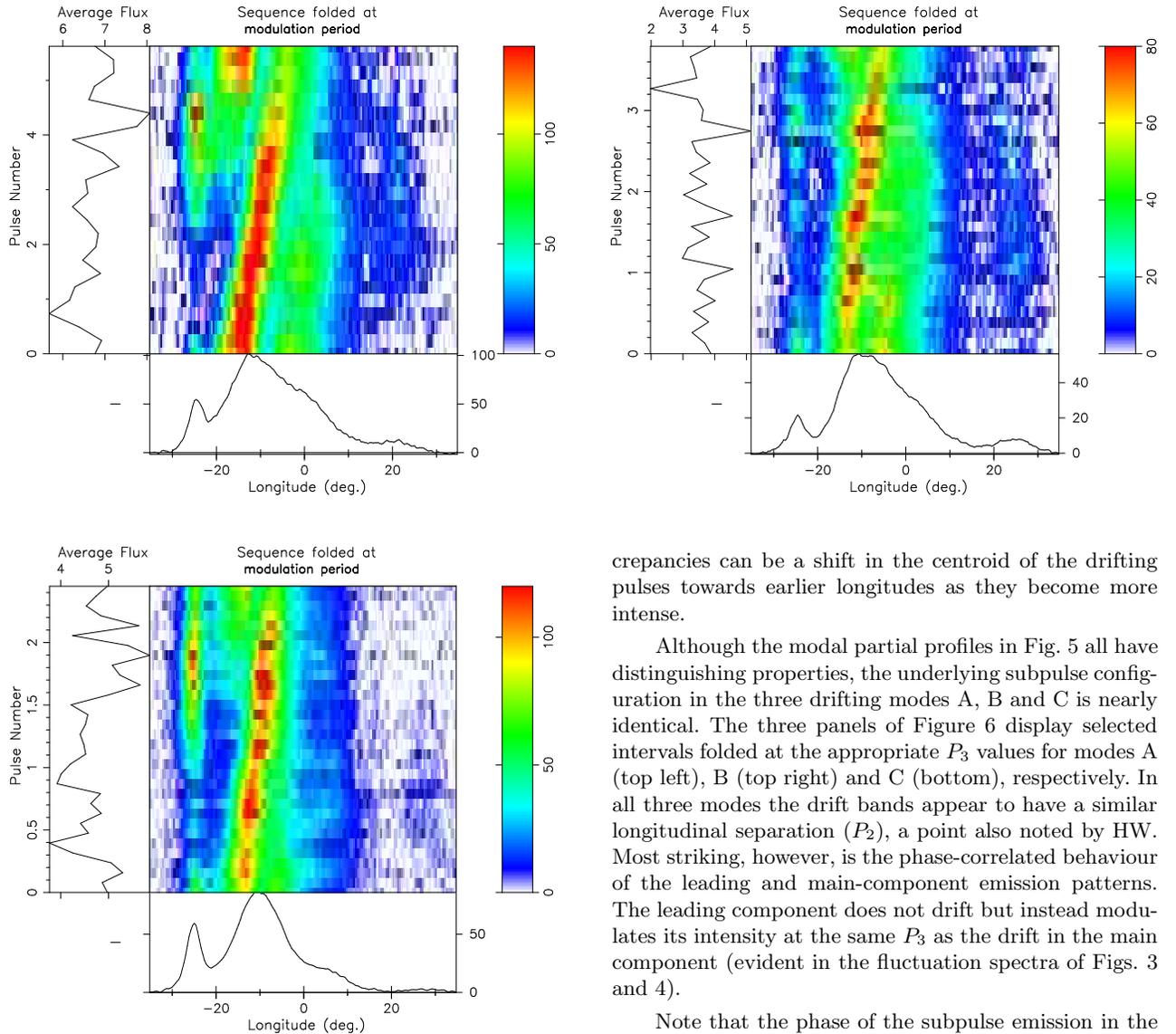}
\caption{Drift-modal sequences folded at their respective $P_{3}$ values: A (top 
left, pulses 2404-2458); B (top right, pulses 2600-2800); and C (bottom, pulses 
481-735). Note the strong similarity of all three patterns, including the relative 
phase positions of the leading and trailing components. Thus, whether aliased 
or not, in each mode the pulsar is executing virtually the same pattern at a different 
speed.}
\label{Fig5}
\end{center}
\end{figure}

By contrast, Fig.~\ref{Fig4} (bottom) shows the same partial profiles, but now 
normalised to the amplitude of the total average profile peak.  This exhibits the 
attenuated latter half of the C-mode profile:  there is no ``hump'' on the trailing 
edge of its main component, and its weak fourth component all but vanishes.  The 
fourth component is strongest in the A mode and appears at a closer longitude 
to the main pulse than the fourth component of the B mode. The plot also enables 
comparison with a similar graph in HW (their fig. 4). The results are broadly compatible, 
but our profiles show a much smaller relative intensity in the region between the 
leading and main profile components, reflecting the much better resolution of our 
observations.  

HW also found that the peak in the N mode falls about 3\degr\ later than in other 
modes, whereas our results do indeed show a peak shift, but only of about 2\degr\ 
(Fig.~\ref{Fig4}, bottom left panel).  A possible explanation for these discrepancies 
can be a shift in the centroid of the drifting pulses towards earlier longitudes as they 
become more intense.

Although the modal partial profiles in Fig.~\ref{Fig4} all have distinguishing 
properties, the underlying subpulse configuration in the three drifting modes 
A, B and C is nearly identical. The three panels of Figure~\ref{Fig5} display 
selected intervals folded at the appropriate $P_3$ values for modes A (top 
left), B (top right) and C (bottom), respectively.  In all three modes the drift 
bands appear to have a similar longitudinal separation ($P_2$), a 
point also noted by HW.  Most striking, however, is the phase-correlated 
behaviour of the leading and main-component emission patterns. The leading 
component does not drift but instead modulates its intensity at the same $P_3$ 
as the drift in the main component (evident in the fluctuation spectra of Figs.~\ref{Fig2} 
and \ref{Fig3}).  

Note that the phase of the subpulse emission in the leading 
component is correlated with the drift patterns in the main components in a 
similar manner in all three modes.  Although the emission level of the leading 
component varies from mode to mode, in each case it is strongest as or just 
before the drift band reappears at the leading edge of the main component. 
A similar correlation is also discernible between the main and trailing components.  
In the A mode (Fig.~\ref{Fig5}, top left) there is evidence of an emission peak 
at the longitude of the trailing component of the main pulse.  Interestingly, 
this peak does not occur at the same phase as that in the leading component, 
but follows it by about 4.5 $P_1$ or 0.65 in phase.  There is also a hint of this 
peak at the same relative phase in the B-mode fold of Fig.~\ref{Fig5}, top 
right, lending an inverse-reflection symmetry to the modulation pattern
[noted also between the two halves of the emission pattern of B0834+06 
(Rankin \& Wright 2007) and in B0818--41, where the phase shift is close to 
0.5 \cite{B09}].

Thus B1918+19 joins the relatively short list of pulsars both classified as having 
double-cone structure and whose single-pulse patterns are phase-locked (B1237+25,  
B0826--34,  B0818--41,  B1039--19:  see Srostlik \& Rankin 2005, Gupta \etal\ 2004, 
Bhattacharyya \etal\  2007, 2011 respectively).  However, as pointed out by Bhattacharyya 
\etal\ (2009), there are as yet no examples where phase-locking is not present in 
double-cone pulsars.  In B1918+19 we can additionally see that the phase-locking 
survives multiple drift-mode changes.  The similarity of the pattern in all three modes 
suggests that the emission processes are simply executing a very similar pattern at 
three different speeds and demonstrates that all components interact as a \emph{single} 
physical system and points to a common geometry and a common emission process.

\begin{figure}
\begin{center}
\includegraphics[height=78mm,angle=-90.]{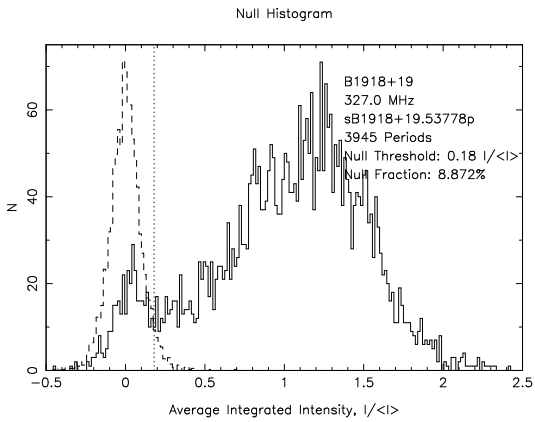}
\caption{Pulse-intensity histogram (solid curve) for the long 327-MHz observation 
of B1918+19 and a corresponding off-pulse noise window (dashed curve).  The 
vertical dotted line indicates a plausible null threshold at 0.18 $<$$I$$>$ (see 
text).  Some 21 pulses were omitted from the distribution due to bad baselines.}  
\label{Fig6}
\end{center}
\end{figure}

\begin{figure}
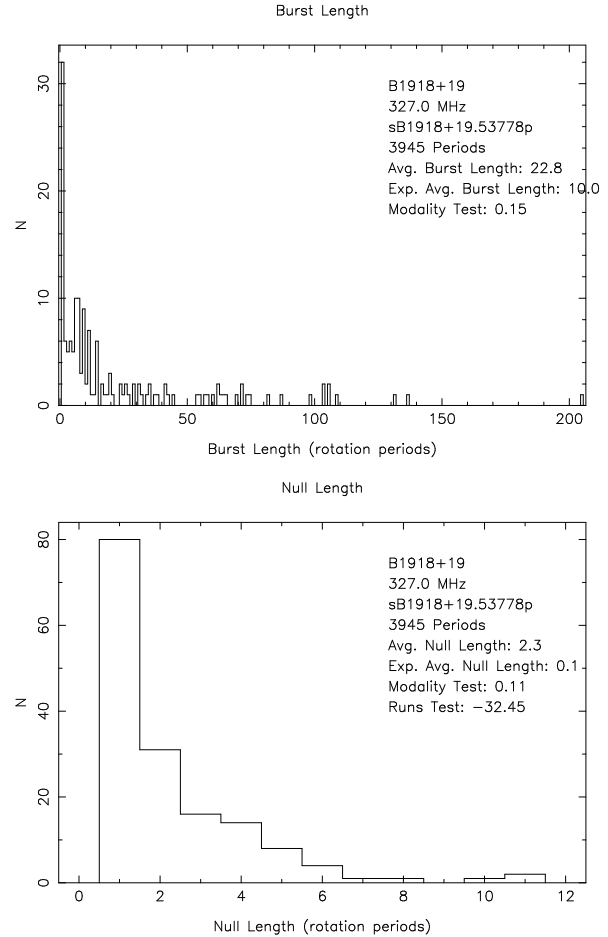

\begin{center}
\includegraphics[height=78mm,angle=-90.]{nulllength_s1918+19_53778p.ps} \\[8pt]
\includegraphics[height=78mm,angle=-90.]{burstlength_sB1918+19_53778p.ps}
\caption{Histograms of null (top) and burst (bottom) lengths.}  
\label{Fig7}
\end{center}
\end{figure}

\section{Quasi-Periodic Null Patterns}
\label{qpnp}
The excellent S/N ratio of the PSs from the upgraded Arecibo instrument enabled 
us to study B1918+19's nulling properties in detail for the first time.  Figure~\ref{Fig6} 
gives the pulse-intensity histogram for the 327-MHz 3946-length PS.  It shows a 
population of weak pulses around zero energy that is distinct from the low-energy 
tail of the primary intensity distribution. This suggested a conservative definition of a 
null as having an intensity less than the minimum in the distribution, which occurred 
at 0.18 $<$$I$$>$.  Using the indicated threshold at 0.18 $<$$I$$>$, the null fraction 
was found to be $\sim$8.9\% or 350 pulses. 

A histogram of null lengths in Figure~\ref{Fig7} (bottom) was computed to investigate 
their statistical properties.  Its most obvious feature is the peak of 80 isolated null 
pulses. If the 350 nulls had appeared randomly among the 3946 pulses with an 
8.9\% chance in each pulse, the expected number of isolated nulls would have been 
some 320. The non-randomness of the distribution is then obvious from the massive 
excess of null pulses occurring in bunches longer than one pulse.  It is also reflected 
in the highly negative ``over-clustered'' Runs-Test value (see Redman \& Rankin 
2008).  

Visual study of the PSs (see Fig.~\ref{Fig1a}) strongly suggests that the nulls are closely 
associated with the N mode, occurring predominantly before, during, after, or within it---so 
it remains possible that nulls are appearing randomly, but confined to this mode of emission.  
If we postulate that all the nulls occur with this set of 847 N-mode pulses---where its 
null fraction would then be 41\% (=350/847)---we could expect some 82 single nulls, 
very close indeed to the number that we observe.  The null-length distribution is also 
statistically overweight in relatively long ($>$3) contiguous nulls. In particular, the 
observed 3 occurrences of 10 and 11 consecutive nulls remain highly improbable. 
It seems an inescapable conclusion that the nulls in this pulsar do have an intrinsic 
bunching tendency such that once a null pulse occurs, the chance of a null in the 
next pulse is enhanced.

We also noted that groups of null pulses often precede and follow a typical 6-10 
pulse burst of emission, even once during the B-mode in an apparent interruption 
of that mode.  In Fig.~\ref{Fig7} (top) a histogram of the emission-burst lengths 
indeed reveals a distinct peak around 6-10 pulses, which can be attributed to this 
emission bunching.  (Also visible is a statistical excess of 1-$P_1$ bursts, reflecting 
emission ``flickering'' within sequences of nulls).  Thus nulls not only tend to occur 
in bunches, but those bunches appear in a quasi-periodic manner---obviously in 
the N mode and often during transitions between modes (see next section). We 
have already noted (\eg, Fig.~\ref{Fig4}) that the N-mode emission is characterised 
by weak leading-component emission and a slight shift in the main peak toward 
later longitudes (see Fig.~\ref{Fig1a}).  These are precisely the properties shared 
by the unexpected 12-$P_1$ feature in the fluctuation spectra of Figs.~\ref{Fig2} 
and \ref{Fig3}, and we are therefore confident that the quasi-periodic nulls associated 
with the N mode are responsible for that feature.

\section{Modal Sequences and Transitions}
Studies of the few pulsars with multiple drift modes (Huguenin \etal\ 1972; 
Wright \& Fowler 1981; Redman \etal\ 2005) have shown that the modes occur 
in sequences of increasing drift rate.  HW found this to be true for pulsar B1918+19 
also, and determined, more specifically, that it exhibits two characteristic modal
sequence patterns: null $\rightarrow$ N $\rightarrow$ A $\rightarrow$ B 
$\rightarrow$ N $\rightarrow$ null, and null $\rightarrow$ C $\rightarrow$ null.  


We found that these two patterns do indeed occur, and they are illustrated in Fig.~\ref{Fig1a}, 
with several examples of the first type shown in the left-hand panel and the C-mode 
dominated pattern on the right. However, they are not the only patterns this pulsar exhibits.  
In particular, we noted frequent examples of the B and N modes being both preceded 
and followed by null bunches (see \S\ref{qpnp}), and there was one possible instance of a 
null $\rightarrow$ B $\rightarrow$ C $\rightarrow$ null sequence. The A mode never seems 
to null and is invariably followed by B mode.  It seems that in all cases a general rule is 
preserved that mode sequences are always from slower to faster {\em observed} drift and are concluded 
by the N mode or nulls. Nevertheless, the C mode, while not violating the rule, tends to 
occur in long PSs with little or no intrusion from other modes, and no examples of A 
$\rightarrow$ B $\rightarrow$ C were found.

In this picture the N mode (together with its accompanying nulls) acts as a kind of reset, 
during which the acceleration of the driftrate ends and the subpulse pattern prepares to 
recommence at a slower rate, be it A, B or sometimes C.  It is accompanied by a fading 
of the leading component and a slight shift of the central intensity to later longitudes.  
Interestingly, we found one ``reset" sequence where the leading component did remain 
strong and gave the emission a periodicity of about 12 $P_1$ punctuated by nulls at this 
periodicity.  This sequence looked more like a fourth slow drift mode than simply a 
disordered N mode and was followed by the A mode, thus being consistent with the 
rule of incrementally increasing driftrates. Interestingly, the role of nulls as agents of 
``reset" is not new: Bhattacharyya \etal\ (2010) noted this effect in B0818--41, a pulsar 
with variable drift, though not with multiple modes.

HW comment that the modal transitions may be more gradual than those found in other 
pulsars (such as B0031--07).   We would suggest that this may not be so.  Initial visual 
inspection of the drift bands seems to indicate that some A $\rightarrow$ B transitions 
may be discrete rather than continuous, while others may actually ``overshoot'' the B 
mode, with the drift rate starting in the slow A mode, then quickly accelerating to even 
become slightly faster than the B mode, only to settle down to the typical B-mode drift 
rate within a few pulses. In the left panel of Fig.~\ref{Fig1a} it is also evident that each 
A $\rightarrow$ B sequence has increasing then diminishing intensity, giving the mode 
change the sense of evolution. This is not seen in the purely C-mode sequence shown 
in the right panel, suggesting that the two types of sequences are of different character, 
with the latter more stable than the first.

\section{The possibility of aliasing}

In the previous section we have analysed only the \emph{observed} drift rate changes 
and, in common with most observers referenced earlier, perhaps tacitly assumed that 
the observed rates are the true rate.  In this section we analyse the consequences of 
challenging this assumption.  

When observing pulsars we sample the emission once every rotation period.  If this 
sampling rate happens to be frequent compared with the timescales of intrinsic changes 
in the emission then we can obtain an adequate description of the phenomenon we 
are seeking to understand.  Assuming this to be true for B1918+19, we may interpret 
the observed drift bands as corresponding to individual subbeams gradually moving 
across the observer's line of sight at the observed drift rate $D$ (the amount of longitude 
phase the subbeam drifts in one pulse).  The repetition rate $P_3$ of this pattern is then 
$P_2/D$ and is significantly larger than $P_1$, the sampling rate. However, from the 
lrf of Fig.~\ref{Fig2}, and above all from the PS in Fig.~\ref{Fig1a},  it is clear that 
$P_3$ switches between several distinct values.  Since $P_2$ remains at roughly the 
same value, the conclusion must be that mode changes arise from changes in the 
subbeam drift rate $D$, which more than doubles as $P_3$ reduces from mode A to 
mode C. 

This is a perfectly self-consistent way of interpreting our observation.  However, any 
physical model must then explain why the subbeams change their intrinsic drift 
speed so drastically, and why these speeds always select from the same values. In the 
framework of the carousel model (Ruderman \& Sutherland 1975; DR; Gil \& Sendyk 2003) 
there is no suggestion as to how this can be achieved and why multiple fixed drift rates 
might be expected.

\begin{figure}
\includegraphics[width=80mm,angle=-90.]{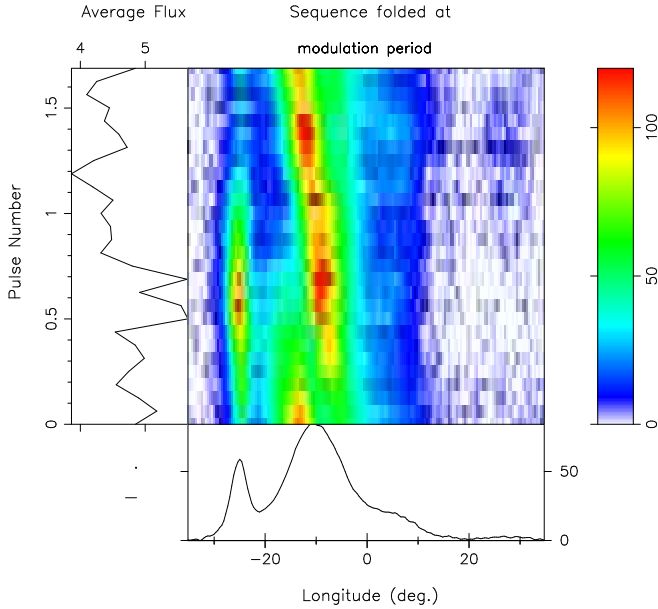}
\caption{Here we see the same C-mode PS of Fig.~\ref{Fig5} (bottom) folded at the 
assumed unaliased $P'_{3}$ of 1/.59 = 1.69 $P_{1}$.   Notice that now the drift is faster 
and from the trailing to the leading edge, and is equivalent to a  fluctuation frequency of 
--0.41 c/$P_1$.  The A- and B-modes can also be folded as in Fig.~\ref{Fig5} with the 
appropriate values of $P'_3$ shown in Table 2, also giving reversed drift directions and 
similarity in pattern.}
\label{Fig11}
\end{figure}

Alternatively, we may assume that our sampling rate is relatively poor, and that we 
sample each subbeam at most twice in successive pulses as it rapidly moves from 
later to earlier phase longitude across our sightline (\ie, first-order aliasing).  This 
means that the apparent drift rate $D$ is no longer a measure of the true drift rate of 
each subbeam, but a confection of the much larger true drift $D'$ and our sampling 
rate.  In other words, between each sampling the subbeams actually drift by an amount 
comparable to the subbeam spacing $P_2$, resulting in an \textit{apparent} drift 
with $P_3$ given by
\begin{equation}
\label{1}
P_3=\frac{P_2}{(P_2-D')}=\frac{P'_3}{(P'_3-1)}
\end{equation}
where $P'_3$ is the true repetition rate and $D'$ is now much greater than the $D$ 
of the non-aliasing scenario.  Now the $P_3$ values measured for B1918+19 (6.06, 3.8 
and 2.45$P_1$) are aliased quantities generated by unaliased $P'_3$ values of 1.12, 1.36 
and 1.69$P_1$, respectively.  Again, the sequences of the modes can be separately folded 
to reveal their patterns, as in Fig.~\ref{Fig5}, but using the new $P'_3$.  An example of 
this is given in Figure~\ref{Fig11}, using an unbroken sequence of C-mode pulses, where 
it can be seen the drift is from trailing to leading edge.

The $P_3$ values are well in excess of the $P'_3$ values which generate them because 
$D'$ and $P_2$ are close in value, yielding a small denominator $(P_2-D')$ in eq. (1).  
Hence the observed, apparently large, reductions in $P_3$ which occur at a mode change 
require only a small change in $D'$, \textit{or a small change in $P_2$}, or both.  Unlike the 
case where $P_3$ is assumed to reflect the true drift, an aliased mode change in $P_3$ 
can be observed even if the drift rate remains unchanged, requiring only a small, perhaps 
observationally imperceptible, change in $P_2$.  This would suggest that a physical account 
of a mode change need not be as drastic as under the non-aliasing assumption.  It may be 
simply the result of somewhat greater or less bunching in the subbeams.  However, the 
question as to why the system always returns to the same discrete modes remains unanswered.

\section{Carousel-model configuration}
The carousel model assumes that the observed subbeams with their corresponding 
$P_2$ and $D'$ reflect the circulation of a number ($n$) of sparks about the pulsar's 
magnetic axis.  Their observed values will depend on frequency and height of the 
emission, but the values of $P_3$, $P'_3$ and $n$ should be frequency and height 
invariant.

In the context of B1918+19, the carousel model places an important geometric/mathematical 
constraint on $P_2$ and $D'$.  If the generating circle of sparks has a circulation time of 
$\P3hat$ pulse periods ($P_1$), then the polar drift rate $D'_{polar}$ is $360\degr/\P3hat$ 
per pulse, and the separation of the sparks is $360\degr/n$, where $n$ has to be 
an integer.  Inserting these into eq.(1) we obtain the intrinsic repetition rate to be
\begin{equation}
\label{2}
P'_3=\frac{\P3hat}{n}
\end{equation}
If we now make the simplest assumption that $\P3hat$ remains constant independently of the mode changes, then we can take the ratios of the mode-dependent values of $P'_3$ derived above to obtain
\begin{equation}
\label{3}
\frac{n_B}{n_A}\approx0.88 \\\frac{n_C}{n_A}\approx0.71
\end{equation}
where $n_A$, $n_B$ and $n_C$ are the number of sparks (subbeams) which will 
produce the appropriate value of $P'_3$ for each mode.  This suggest a self-consistent 
model with integral values of $n$ where $n_A=10$, $n_B=9$ and $n_C=7$, and implies 
from eq.(2) a near-constant circulation time $\P3hat$ between 11.8 (from C) and 12.2 
(from B), with $n_A=10$ giving $\P3hat=12.0$ exactly  (a model with integral multiples 
of these values for $n$ is theoretically valid, but unlikely to have physical meaning).

What consequences does the assumption of a fixed circulation time of 12-$P_1$ have 
on our understanding of the observations? From eq.(1) the observed aliased $P_3$, 
caught by the eye and measured in the lrf, is (in units of $P_1$)
\begin{equation}
\label{4}
P_3=\frac{\P3hat}{(\P3hat-n)}
\end{equation}
Note that for values of $n$ close to $\P3hat=12$, $P_3$ will be large (giving, for example, 
$P_3=6$ for $n=10$ in the A mode) but decrease with decreasing $n$.  Thus we arrive at 
an interesting and self-consistent picture where the A mode commences with 10 subbeams, 
then loses a beam and rapidly adjusts to a B  mode with 9 sparks with a slightly enlarged 
spacing, sometimes progressing to a C mode of 7 subbeams.  The spacings of the A mode 
beams will be just 70\% smaller than those of the C  mode.  However, the resulting changes 
in $P_2$ may be difficult to measure in practice (see next \S) but could be a test of the model. 

Notice that this is almost precisely what can be seen in Fig.~\ref{Fig3}, where fluctuations 
peaking near 0.83, 0.75, and 0.58 c/$P_1$ are evident.  One can even see vestiges of a 
weak periodicity at at 0.67 c/$P_1$ that corresponds to a carousel with 8 subbeams, and 
although we have never identified any stable PS with these properties, it does seem to be 
present as a transitional structure when the A mode sometimes ``overshoots'' the B and then 
resets to the B.  

The role of the N mode is interesting. We have noted earlier that this mode is accompanied 
by a periodicity with $P_3\approx12 P_1$, partly generated by repeated null bunches embedded 
in the N mode.  Can it be a coincidence that this is precisely the value deduced by separate 
arguments for the fixed circulation period?  There are two ways this might happen.  Firstly, 
according to eq.(4) this value would be consistent with a circulation of $\P3hat\approx12 
P_1$---but with exactly $n=11$ subbeams.  Thus the sequence $N\rightarrow{A}\rightarrow{B}\rightarrow{N}$ corresponds to a one-by-one reduction in subbeams from 11 to 9 and re-set.  Extrapolating 
further, 12 subbeams would, through aliasing, formally produce an infinite $P_3$---\ie, 
sustained non-drifting emission with a narrow $P_2$ during the relatively short duration 
of the often irregular N  mode.  Thus the sequence may even be argued as starting at 
$n=12$ (see Table 2).

A second---or additional---way of explaining the N-mode's periodicity would be to speculate 
that the subbeam system breaks up as $n$ reduces below the C-mode value of 7.  At low 
values of $n$ the subbeam separation $P_2$ would rapidly grow to twice or more than the 
values in the A, B and C modes (see Table 2), so that when $n$ equals, say, 3 or 4 the 
observer would see a flickering or a series of pulses with deep nulls between the subbeams, 
depending on the width of the subbeams.  Once the system has degenerated, the circulating 
system might just be partly null and partly irregularly-located beams, so that in the N mode 
we see the circulation time directly.  Thus the N mode could be explained as both being the 
ending and the beginning of the modal cycles, emphasizing its role as the mode during which 
the cycles are reset. 

Obviously missing in the sequence is the case of $n=8$ subbeams, where a $P_3$ of 3.0 
$P_1$ would be expected.  However, we note that there is no clear evidence of the B mode 
progressing directly to the C mode, suggesting that the C mode arises under separate 
circumstances and is observed to be steadier than the A and B modes (Fig. 2).  The reason 
for the stability of the C mode is suggested by the form of eq.(4): when $n$ is close to 
$\P3hat\approx12 P_1$, as it is in modes A and B, small modulations in $\P3hat$ may 
cause observable modulations in $P_3$, but when n is smaller (7), as in mode C, 
modulations in the drift rate will not significantly modulate the aliased $P_3$. 

A summary of this picture is given in Table 2.  The observed values of $P_3$, assumed 
to be first-order aliased, are shown unbracketed, together with their inferred true drift 
repetition rates $P'_3$ and inferred circulation times $\P3hat$.  The near-invariant 
circulation time of 12 $P_1$ is then assumed to deduce the expected $P_3$ and $P'_3$ 
for other integral values of $n$, and these are shown in brackets. In the last column we 
give the angular spark separation for each $n$.  Note how these increase weakly for 
high $n$ but widen dramatically for small $n$.   

\begin{table}
\begin{center}
\caption{Carousel model expectations of the circulation time ($\P3hat$) for different n assuming $P_3$ is the first alias of $P'_3$. Comparisons corresponding to observed $P_3$ values are shown bold and  unbracketed.}
 \begin{tabular}{cccccc}
 \hline
 \hline
$n$ & $P_3$ & $P'_3$ & $\P3hat$ & $360/n$\\
  & ($P_1$) &  ($P_1$) & ($P_1$) & (\degr)\\
 \hline
 12 & ($\infty$) & (1.00) & (12.00) & 30\\
 \bf{11} & \bf{12.0} & \bf{1.09} & \bf{12.00} & \bf{32.8}\\
 \bf{10} & \bf{6.06} & \bf{1.20} & \bf{12.00} & \bf{36}\\
 \bf{9} & \bf{3.8} & \bf{1.36} & \bf{12.24} &  \bf{40}\\ 
 8 & (3.0) & (1.50) & (12.00) & 42.5\\
 \bf{7} & \bf{2.45} & \bf{1.69} & \bf{11.83} & \bf{51.5}\\
 6 & (2.00) & (2.00) & (12.00) & 60\\
 3 & (1.33) & (4.00) & (12.00) & 120\\
\hline
\end{tabular}
\end{center}
\label{Table2}
\end{table}

Finally, if B1918+19's circulation time $\P3hat$ is about 12-$P_1$, and the N, A, B and 
C modes reflect carousel configurations with 11, 10, 9 and 7 subbeams, respectively, then 
how are we to understand the 85-$P_1$ fluctuation feature found in the lrf  of in Fig.~\ref{Fig2}?  
As mentioned above, this feature is null-associated, and while from the hrf of Fig.~\ref{Fig3} 
it seems to be a combination of amplitude and phase modulation, it is carried by such 
small Fourier components that it must be a modulation of the entire profile.  Interesting 
also is the fact that one sees little of this feature in the analysis of shorter PSs.  The 
A-mode PSs are too short, and those in the C mode do not seem to null.  Rather it is 
the B and N modes that are typically punctuated by nulls, and the former are usually 
rather long, averaging 53 pulses, while the N mode averages 23.5 pulses.  

What we therefore believe, and what is evident from our PS in the left panel of Fig.~\ref{Fig1a}, 
is that this feature is associated with the full modal cycle of N$\rightarrow$A$\rightarrow$B$\rightarrow$N, 
which tends to start and end with nulls. Similarly,  the 43-$P_1$ feature arises from the 
shorter and less frequent N$\rightarrow$B$\rightarrow$N cycle.  The cycle involving the 
C mode is of a different character and never exhibits this feature.



\section{Emission geometry}
In the previous section it has been argued on purely mathematical grounds that the 
apparent repetition patterns of B1918+19's emission can arise from first-order aliasing 
of equally-spaced circulating subbeams.  Assuming the circulation rate to be unvarying, 
it was possible to show that each of the observed drift rates corresponded to a different 
integral number of subbeams, and hence to a specific angular spacing of those subbeams 
(Table 2).   

We must now ask whether the angular spacing generated at the pulsar's polar cap 
for each mode is compatible with the spacing $P_2$ of the emission beams at our 
observing frequency. This spacing is only relevant for drifting subpulses, implying 
here that it must be evaluated for the inner cone in the central part of the profile, not 
the outer conal components on the profile edges.  And, of course, we are interested 
in the values for each of the three drifting modes, A, B and C.

\begin{figure}
\includegraphics[height=80mm,angle=-90.]{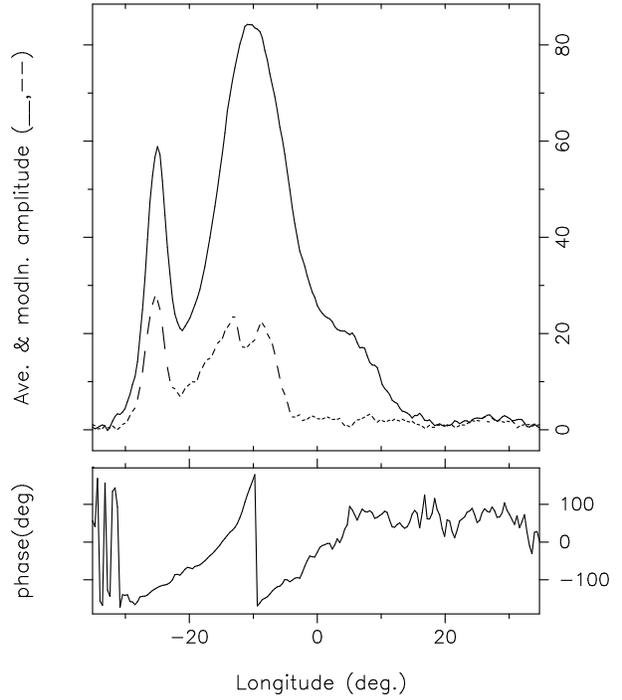}
\caption{Modulation phase diagram for the C-mode PS in Fig.~\ref{Fig11}.}
\label{Fig12}
\end{figure}

Measuring $P_2$ accurately is never trivial, and B1918+19 entails several further 
complications:  first, the always short A-mode episodes introduce one kind of 
difficulty and the incomplete trailing part of the main component in the C mode 
another.  With 20/20 hindsight we can see that HW's $P_2$ values in their table 
are inconsistent, and we will argue that they are least correct with their C-mode 
value.  HW determined their values directly by measuring the intervals between 
subpulses.  Another method uses the phase ramp of the lrf spectra, and we give 
such a diagram for the C mode in Figure~\ref{Fig12}.  There it can be seen both that 
the fluctuation power is virtually absent under the trailing part of the profile, but 
in the immediately adjacent leading part, the slope is such that a 360\degr\ 
excursion would correspond to about 18-20\degr\ of longitude.  This gives an 
estimate of the C-mode $P_2$, but a poor one owing to the restricted longitude 
range of the modulation.  Similar phase diagrams for short B- and A-mode PSs 
(pulses 2459-2520 and the 55 A-mode pulses in Fig.~\ref{Fig5}) provide better 
estimates, suggesting $P_2$ values of some 14-15\degr\ and 13$\pm$2\degr, 
corresponding to 33 and 30$\pm$5 ms, respectively (40$\pm$4 and 33$\pm$3 
ms in HW).  

\begin{table}
\begin{center}
\caption{Geometrically deduced values of n for modes A, B and C.}
 \begin{tabular}{cccccc}
 \hline
 \hline
$\alpha$ & $\beta$ & $P_2$ & $\eta$ & $n$\\
 (\degr) &(\degr)  & (\degr) &(\degr) &      \\
 \hline
 13.7& --4.2 & 13 & 36.1 & \bf{10.0} \\
        . &      .    & 14.5 & 40.4 & \bf{8.9} \\
         . &      .    & (16)& (44.7) & (8.0) \\
         . &       .   &  18 & 50.6 &  \bf{7.0} \\ 
\hline
\end{tabular}
\end{center}
\label{Table3}
\end{table}

\begin{figure}
\includegraphics[height =80mm,angle=-90.]{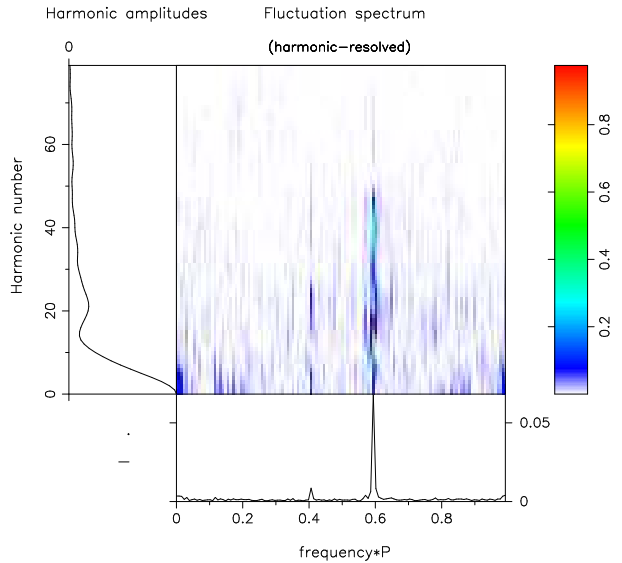}
\caption{The hrf for the C-mode PS in Fig.~\ref{Fig5}.   Note that the drift modulation 
is carried by Fourier components around or just greater than 20 and that a second 
harmonic is discernible.  These imply that the $P_2$ for this mode is 
$\approx360\degr/20=18\degr$.}
\label{Fig10}
\end{figure}

$P_2$ values can also be estimated from the hrf spectra, and we computed an 
hrf with the longitude restricted to the central component (not shown).  This did 
not seem to significantly affect the three features corresponding to the drift modes.   
However, taking the hrf of the relatively long C mode PS (Figure~\ref{Fig10}) we 
can see that the primary modulation is associated with the Fourier component 
peaking at about 20 (its harmonic very clearly peaks about component 40); 
therefore, $P_2$ here is very near 360\degr/20 or 18\degr\ or 41 ms (34$\pm$2 ms 
in HW).  We found it difficult to measure the A- and B-mode $P_2$ values by this 
method, and the problem is discernible in Fig.~\ref{Fig3}.  The harmonic numbers 
carrying the B and A modes are clearly smaller as can be judged from their 
harmonics, but the peaks themselves are too broadened and conflated to give 
any reliable estimate.

The above $P_2$ values are given in Table 3, and we can use these values 
for each mode, together with the inferred geometry of the pulsar with respect to 
us, to compute what kind of carousel subbeam system underlies them.  Starting 
with the profile geometry in \S3 above, it is of importance here is that the deduced 
$\alpha$ and $\beta$ values square with the central PPA rate 
$R$ [$=\sin{\alpha}/\sin{\beta}$] of --3.2\degr/\degr.  The circumstance that this 
pulsar has an inside (poleward) sightline traverse (as does B0943+10) is also quite 
clear from the profile geometry and linear character of the star's PPA traverse.  

The point is that $P_2$ subtends a particular magnetic polar 
longitude $\eta$, which in turn can be used to estimate the subbeam number.  
DR give such a relation as their eq. (3), but Gupta \etal's eq. (6) is more accurate 
because it distinguishes the conal edges from the subbeam path within the cone.  
We used a version of this relation $\sin\eta=\sin(\alpha+\beta)\sin P_2/\sin\Gamma$ 
where the subbeam path $\eta$ is 0.8\degr\ interior to the outside conal half power 
point.  Note that this value closely reflects the difference between the outer half-power 
and peak conal radii in the models of ET VI and Gil \& Sendyk (2000).  Therefore, in 
our case, $\beta$ is negative, so the peak subbeam track has a magnetic polar 
colatitude of $|\beta + 0.8\degr|$ or 3.4\degr.  Then, the subbeam number $n$ can 
be estimated as 360\degr/$n$.  These simple computations are assembled in 
Table 3, where it can be confirmed that the derived subbeam numbers are near 
integral and are fully compatible with the integers 10, 9 and 7 for the A, B and C 
configurations respectively in Table 2.  As a check, one can also carry out these 
computations for a positive (equatorward) sightline geometry, and the resultng 
$\eta$ values are roughly twice as large, implying a very small number of subbeams, 
demonstrating that this sightline would be incorrect.  


\section{Summary of results}
In 1987 Hankins \& Wolszczan carried out the first analysis of the properties of B1918+19, 
attracted by its entirely conal triple profile and the added opportunities this configuration 
presents to investigate the relationships of emission in its inner and outer cones.  Using 
Arecibo PSs they were able to identify the four basic single-pulse modes of emission 
(A, B, C and N) and noted specific cyclic relationships between these modes.

Here, with the advantage of more recent data from the same telescope but at a slightly 
lower observing frequency, we explore the geometry of the pulsar and refine and add 
to the known single pulse repetition patterns.  Further, we propose a possible carousel 
configuration present at the polar cap which may explain the mode changes with a simple 
idea. Before discussing this model in the next section, it is useful to summarise our 
observational results: 

\begin{itemize}

\item B1918+19 has integrated properties common to many pulsars: it has a conal 
triple (cT) profile with very nearly the usual angular 	dimensions for its inner and 
outer cones---half-power conal radii $\rho$ at 1-GHz of 4.1 and 6.5\degr---as 
computed originally in ET VI.  Its integrated polarisation properties suggest that 
the magnetic latitude $\alpha$ and sightline impact angle $\beta$ are some 13.7 and 
--4.2\degr, respectively, such that its inside (poleward) sightline traverse cuts its inner 
cone tangentially ($\beta/\rho\sim -0.90$) and outer  cone obliquely ($\beta/\rho\sim -0.65$)

\item The profile exhibits leading/trailing asymmetry of two types:  the leading (outer) 
conal component is much stronger than the trailing one; and the leading part of the 
main (inner cone) component is stronger than the trailing part.  This asymmetry is 
least marked in the N mode, and becomes stronger in the A and B as well as the C 
mode---wherein trailing portions of the profile nearly vanishes.

\item As HW found, the star's pulse sequences exhibit three highly discrete drifting 
modes, A, B and C, as well as a non-drifting N mode, However, the pulsar exhibits 
\textit{five} prominent fluctuation features, three corresponding to the A, B and 
C drift modes with $P_3$ values near 6.0, 4.0 and 2.5 $P_1$, and two null-related 
features at about 12 and 85 $P_1$. 
 
\item About 9\% of the pulses are ``nulls'' but these nulls are distributed in a highly 
non-random manner, evidenced by the substantial fraction of nulls with long 
(2-10 pulse) durations.  The A and C modes seem not to null, and the B mode 
infrequently, so the nulls occur mainly within the N mode at a roughly 43\% 
fraction.

\item Fluctuations in the inner-cone main and outer-cone leading components are 
phase-locked in all three modes---and with very similar phase relationships.  This 
suggests that the drift features, A, B and C, may be seen as consisting of a similar 
pattern across both inner and outer cones spun past our sightline at three different 
speeds, or as a widening pattern at a common speed. The phase-locking is very 
reminiscent of the double-cone model proposed for B0818--41 by Bhattacharyya \etal\ 
(2007), and has been argued for B1237+25 (Srostlik \& Rankin 2005), B0836--24 
(Gupta \etal\ 2004, Esamdin \etal\ 2005) and B1039-19 (Bhattacharyya \etal\ 2011). 
Finding this same property in B1918+19 mode-by-mode suggests that locking between 
inner and outer cones is a common, maybe ubiquitous, feature in double-cone pulsars 
with regular drift. 

\end{itemize}

\section{Concluding Remarks}
For B1918+19 the classic carousel model of Ruderman \& Sutherland (1975) predicts 
a circulation time $\P3hat$  of 8.5 $P_1$, well below what would be expected if the 
observed $P_3$ values were interpreted directly as measures of spark-to-spark 
transition speeds. With a more subtle model involving a central discharge (Gil \& 
Sendyk 2000) it may well be possible to achieve a more plausible circulation speed, 
but it remains hard to understand why the circulation time switches between three 
fixed but very different values in this and other pulsars (see, for example, discussion 
of the quantised drift patterns of B2319+60 in Wright \& Fowler 1981).

Here we have explored the view that the observed driftbands result from first-order 
aliasing of a much faster drift. This is what the original carousel model (assuming 
any reasonable number of ``sparks") would actually \textit{predict}, since our 
sampling rate ($P_1$) is comparable to the frequency ($P'_3$) with which the 
"sparks" are presented to us\footnote{Note that higher orders of aliasing are 
perfectly possible in mathematical terms. These will require even faster circulation 
times and have not been preferred here for physical reasons.  However, they 
cannot be excluded for different physical configurations and in carousels of other 
pulsars.}.  This is not the first time subpulse drift has been ascribed to aliasing 
(\eg, DR, Janssen \& van Leeuwen 2004, Gupta \etal\ 2004, Bhattacharya \etal\ 2007), 
but here we extend the idea to account for the presence of quantised drift rates and 
their sequence. Using our observations we are able to \textit{deduce} that aliasing 
requires 10, 9 and 7 subbeams corresponding to the A, B and C modes and, more 
surprisingly, that these are all consistent with a common circulation time of about 
12 $P_1$, not far from the original RS figure. This essentially mathematical-geometric 
picture appears to be compatible with the observed $P_2$ values, which increase 
gradually as $P_3$ falls, as the model requires (see Table 2).  Thus we no longer need 
to explain a changing circulation time, and can understand the changing---quantised---drift 
rates as a progressive reduction in the number of sparks.  This is physically much easier 
to understand.

It is significant that the N mode is also clearly linked to the circulation time. The N 
mode was not involved in the deduction of A, B and C's common $\P3hat$, and 
yet its repetition rate $P_3\approx12$ -- discovered here -- is virtually the same 
as $\P3hat$. So what does this mode represent? We demonstrate in \S 9 that the 
mode can betray the underlying circulation time by firstly ending the mode cycle 
through the vanishing subbeams with wide angular separations, and then by re-starting the 
mode cycle through the aliasing of 11 subbeams. Thus the mode sees the 
regeneration of the mode cycle and must be the phase in that cycle when the 
subbeams re-establish themselves.


One may speculate whether carousel configurations with varying numbers of 
subbeams might explain the multiple different drift modes exhibited by other pulsars. 
The best-known examples, B0031--07 (Huguenin \etal\ 1970; Vivekanand \& Joshi 
1997), B2319+60 (Wright \& Fowler 1981), B2303+30 (Redman \etal\ 2005) and 
B1944+17 (Deich \etal\ 1986, Kloumann \& Rankin 2010), have similar ratios in 
their  $P_3$ values to those of B1918+19 and are worthy of further investigation in 
the context of aliased drift.

However, as is evident in the sequences of Fig. 2 and as noted above, the drift rates 
of B1918+19 are not totally steady in the A and B modes and at the transition can 
initially overshoot their new value. This suggests that the circulation time is not held 
rigidly steady as supposed in the basic calculations here. Pulsars B2016+28, 
B0818--41 and B0826--34 all intermittently display curved driftbands, so changes in 
pulsar drift patterns cannot always be explained by the sudden disappearance of a 
subbeam. 

The point we make in this paper is that subtle variations in the circulation time and 
in the number of circulating beams can, if the sampling rate of once per pulse period 
is infrequent for probing the true circulation time, generate the extraordinarily dramatic 
and spectacular effects seen in the pulse sequences of many pulsars.

\section*{Acknowledgments} GAEW is grateful to the Astronomy Centre, University of Sussex, for a Visiting Research Fellowship. Much of the work was made possible by visitor-grant support from the US National Science Foundation Grants 00-98685 and 08-0769.  Arecibo Observatory is operated by SRI International under a cooperative agreement with the National Science Foundation, and in alliance with Ana G. M\'endez-Universidad Metropolitana, and the Universities Space Research Association..  This work made use of the NASA ADS astronomical data system.

{}

\label{lastpage}

\end{document}